\documentclass{sf2a-conf2014}
\usepackage{graphicx}
\usepackage{hyperref}
\usepackage[]{natbib}  
\usepackage{epstopdf}

\def\BibTeX{{\rm B\kern-.05em{\sc i\kern-.025em b}\kern-.08em
    T\kern-.1667em\lower.7ex\hbox{E}\kern-.125emX}}
\bibpunct{(}{)}{;}{a}{}{,}  

\begin{document}

\TitreGlobal{SF2A 2014}


\title{Influence of different parameters on the chemical composition of warm Neptunes}

\runningtitle{Influence of different parameters on the chemical composition of warm Neptunes}

\author{O. Venot}\address{Instituut voor Sterrenkunde, Katholieke Universiteit Leuven, Celestijnenlaan 200D, 3001 Leuven, Belgium}

\author{M. Ag\'undez}\address{Instituto de Ciencia de Materiales de Madrid, CSIC, C/Sor Juana In\'es de la Cruz 3, 28049 Cantoblanco, Spain}

\author{F. Selsis$^{3,}$}\address{Univ. Bordeaux, LAB, UMR 5804, F-33270, Floirac, France}\address{CNRS, LAB, UMR 5804, F-33270, Floirac, France}
\author{M. Tessenyi}\address{University College London, Department of Physics and Astronomy, Gower Street, London WC1E 6BT, UK}
\author{L. Decin$^{1}$}




\setcounter{page}{237}


\maketitle


\begin{abstract}
We developed a 1D photo-thermochemical model to study the atmosphere of warm exoplanets. The chemical scheme used in this model is completely new in planetology and has been constructed in collaboration with specialists of combustion. It has been validated as a whole through experiments on a large range of temperature (300 - 2500 K) and pressure (1 mbar - 100 bar), allowing to study a wide variety of exoplanets.
We have used this chemical model to study the atmosphere of two warm Neptunes, GJÊ3470b and GJÊ436b, and the influence of different parameters (vertical mixing, metallicity, temperature, \dots) on their chemical composition. We present here the results obtained in these studies.
\end{abstract}



\section{Introduction}

Transit spectra have recently allowed us to characterise the atmosphere of the Neptune-sized planet like GJ~436b, GJ~3470b, or the super-Earth GJ~1214b, which orbit around M dwarf stars. These planets present interesting differences with respect to hot Jupiters. The first is because the host M dwarf star is smaller and cooler than a solar-type star, so the planet is less severely heated (even if the orbital distances in the range 0.01 -- 0.04 AU are comparable to the ones of hot Jupiters). This results in planetary effective temperatures below 1000~K. Interestingly, it is around this temperature that a gaseous mixture at thermochemical equilibrium, with solar elemental abundances and at a pressure around 1 bar, shows a transition concerning the major carbon reservoir, either CO or CH$_4$ for temperatures above and below 1000~K, respectively \citep[see][their Fig. 1]{venot2014}. In this regard it is interesting to note that transit spectra of GJ~436b indicates that its atmosphere is dominated by CO \citep{ste2010, mad2011, knu2011}, whereas methane is predicted to be the major carbon reservoir at thermochemical equilibrium. However, such interpretation has been disputed by \cite{bea2011} based on a different analysis of transmission spectra. From a modelling point of view, \cite{line2011}, with a code considering thermochemical kinetics, vertical mixing, and photochemistry, concluded that CH$_4$ should be the major carbon-bearing molecule in GJ~436b's atmosphere.

A second important difference with respect to hot Jupiters is that (sub-)Neptune planets have a lower mass. Thus, they have a lower efficiency to retain light elements \citep{elk2008}, so we can expect their atmosphere to be enriched in heavy elements with respect to the solar composition.
Previous photochemical studies dedicated to (sub-)Neptunes  explored metallicities reasonably high (up to 50 $\times$ solar metallicity for \citealt{line2011, mil2012}) but also extremely high (up to 10,000 $\times$ solar metallicity for \citealt{mos2013}).
In this study, we considered an enrichment in heavy elements between 1 and 100 $\times$ solar. Enrichment in heavy elements in the range 50-100 is extremely interesting because it corresponds to a change of the carbon reservoir (either CH$_4$ or CO) for pressures within 1 and 100 bar and temperatures within 1000 and 2000~K \citep[see][their Fig. 1]{venot2014}. The deep atmospheric layers, with such high temperatures and pressures, can then contaminate most of the atmosphere due to the chemical quenching associated with vertical mixing \citep[e.g.][]{prinn1977carbon, lewis1984vertical, vis2011, moses2011, venot2012}.
 
\section{Chemical scheme}

The first studies on hot Jupiters \citep{liang2003, liang2004} used chemical schemes made originally for Jupiter or Saturn, so cold atmospheres in which endothermic processes happening only at high temperature were not included. Then, some improvements have been done, by adding exothermic reactions to these chemical schemes \citep[i.e.][]{zahnle2009a, zahnle2009b, line2010, moses2011}, but none of these studies discuss about the validation of the chemical network.  In order to study correctly these hot exoplanets, we have developed a chemical scheme adapted to the high temperatures and pressures that are found in warm exoplanet atmospheres. Because such extreme conditions are also found in car engines, we have collaborated with specialists of combustion, working mainly for industrial purposes. We have implemented a chemical scheme new in planetology, that has been validated experimentally as a whole and not only for each individual reaction as it is often the case for chemical schemes used in planetology. This validation has been performed on a wide range of pressure and temperature (0.01 - 100 bar and 300-2500~K), allowing to explore a variety of atmospheric environments. The scheme includes 957 reversible and 6 irreversible reactions (so a total of 1920 reactions), involving 105 neutral species (molecule or radical). Helium is included in this mechanism and plays the role of third body in some reactions. It is available in the online database KIDA: KInetic Database for Astrochemistry\footnote{\url{http://kida.obs.u-bordeaux1.fr}} \citep{wakelam_kida2012}. More details about this chemical scheme and its experimental validation can be found in \cite{venot2012}.

Data used to model photolysis processes need also to be improved. Indeed, absorption cross sections, which are crucial to calculate photodissociation rates, are known only at ambiant and low temperature. In this view, we have measured the absorption cross section of CO$_2$ in the wavelength range [115-230] nm at temperatures up to 800K \citep{venot2013}. Measurements of other molecules are now needed, and are currently in progress.

\section{GJ 3470b}

GJ 3470b is a transiting warm Neptune discovered by \cite{bon2012}. We used this planet as study-case to investigate the effect of temperature, vertical mixing, metallicity and UV irradiation on its atmospheric chemical composition. We explored the parameter space as explained in Table \ref{tab:parameterGJ3470b}. The 17 models we have computed allowed us to frame the different compositions that are possible for this planet.

We were interested in the CH$_4$/CO ratio, which is a parameter quite controversial. Indeed, the identification of the C-bearing species from observations is still under debate for warm Neptunes such as GJ~436b \citep{ste2010, mad2011, bea2011}. The results we obtained in this study show that the situation is also not simple from a chemical point of view. CH$_4$ may or may not be the major carbon reservoir, depending on the metallicity, the temperature, and the vertical mixing. Indeed, we found that in most of cases the CH$_4$/CO ratio is above 1, but CO can be more abundant in the case of a high metallicity ($>$~100~$\times$~solar) combined with an atmospheric temperature high too. These conditions are plausible, because an enrichment with respect to the solar elemental abundances can be expected for planets with a similar bulk composition as Uranus and Neptune. Because of similar physical properties, this result can be extrapolated to other warm (sub-)Neptunes, such as GJ~436b or GJ~1214b. Recently, a similar study has been carried out by \cite{mos2013}, who also find that a high metallicity could lead to a CH$_4$/CO ratio lower than 1 in GJ~436b. A very high metallicity (100 $\times$ solar metallicity) seems to be a solution to explore in order to interpret future observations, as it is very likely for these atmospheres.

The synthetic spectra we computed (see \citealt[their Figs. 7 and 8]{venot2014}) indicate that the brightness temperature, as well as the transit depth, vary significantly with the metallicity and the thermal profile, so future observations of GJ 3470b may be able to determine the metallicity and the temperature of this planet. Because of the strong opacities, spectra corresponding to high metallicity models (100~$\times$~solar) produce smaller features than low metallicity models (1 $\times$ solar). On the spectra corresponding to the primary transit, we found that the 3.3-to-4.7 $\mu$m ratio changes together with the CO/CH$_4$ ratio. Observations at these wavelengths are a possible way to constrain this ratio.

These results are presented in more details in \cite{venot2014}.

\begin{table}
\centering
\begin{tabular}{llll}
\hline \hline
Parameter      & Range of values                                         & Symbol \\
\hline
Metallicity     & Solar ($\zeta=1$)       & $\zeta_1$ \\
                     & High ($\zeta = 100$) & $\zeta_{100}$ \\
\hline
Temperature & Warm atmosphere ($+100$~K)        & $T_{+100}$ \\
                     & Cool atmosphere ($-$100~K)       & $T_{-100}$ \\
\hline
Eddy diffusion coefficient & High ($K_{zz}$ $\times 10$)     & $K_{zz}^{\times 10}$ \\
                                         & Low ($K_{zz}$ $\div 10$)           & $K_{zz}^{\div 10}$ \\
\hline
Stellar UV flux                   & High irradiation ($F_{\lambda}$ $\times 10$) & $F_{\lambda}^{\times 10}$ \\
                                         & Low irradiation ($F_{\lambda}$ $\div 10$)       & $F_{\lambda}^{\div 10}$ \\
\hline
\end{tabular}
\caption{The parameter space of the model explored. All the parameters are changed with respect to the standard values showed in \cite{venot2014} (their Figs. 3 and 4). The standard metallicity is 10 $\times$ solar ($\zeta = 10$).} \label{tab:parameterGJ3470b}

\end{table}

\begin{figure}[ht!]
 \centering
 \includegraphics[width=0.8\textwidth,clip]{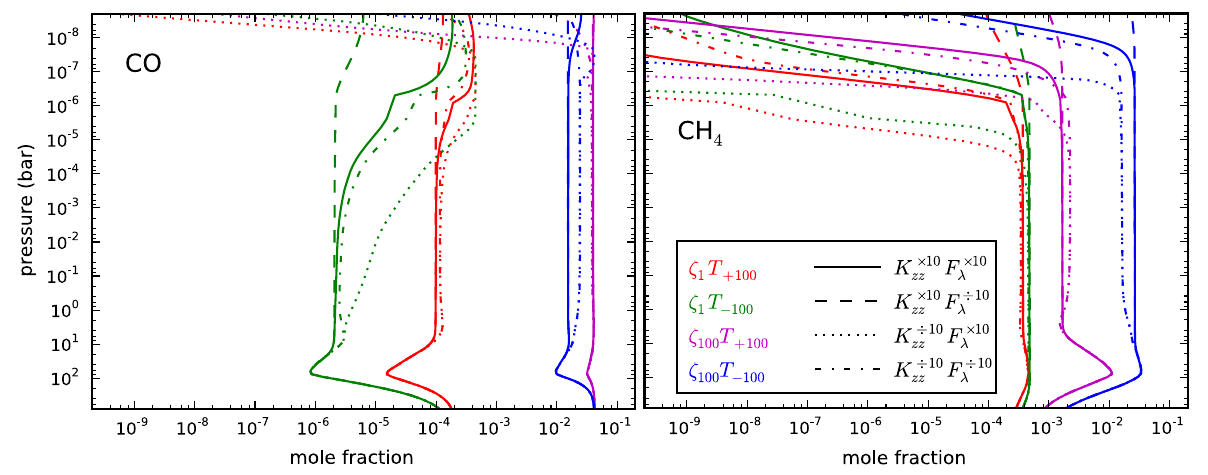}
  \caption{Vertical abundance profiles of CO and CH$_4$ as calculated through each of the 16 models in which the space of metallicity,
temperature, eddy diffusion coefficient, and stellar UV flux are explored. Each colour corresponds to a set of metallicity and temperature values
and each line style to a set of eddy diffusion coefficients and stellar UV fluxes (see legend in the CH$_4$ panel and meaning of each symbol in
Table \ref{tab:parameterGJ3470b}.}\label{CO_CH4_GJ3470b}
\end{figure}

\section{GJ 436b}

We did a study quite similar on GJ 436b, which is a warm Neptune with a high eccentricity (e~=~0.16). Thus, the planet undergoes strong tidal forces. The dissipation of this energy in the planet depends on its internal composition and structure, which are both unknown. We have used a Constant Time Lag model to determine the range of internal temperatures that are possible for this planet. We found that the maximum internal temperature possible was 560 K. In our modelling, we considered 4 internal temperatures: 100, 240, 400, and 560 K. We have also explored the sensitivity of the atmospheric chemical composition to the metallicity by considering 3 cases: $\zeta$ = 1, 10, and 100 $\times$ solar metallicity.

\begin{figure}[ht!]
 \centering
 \includegraphics[width=0.345\textwidth,clip]{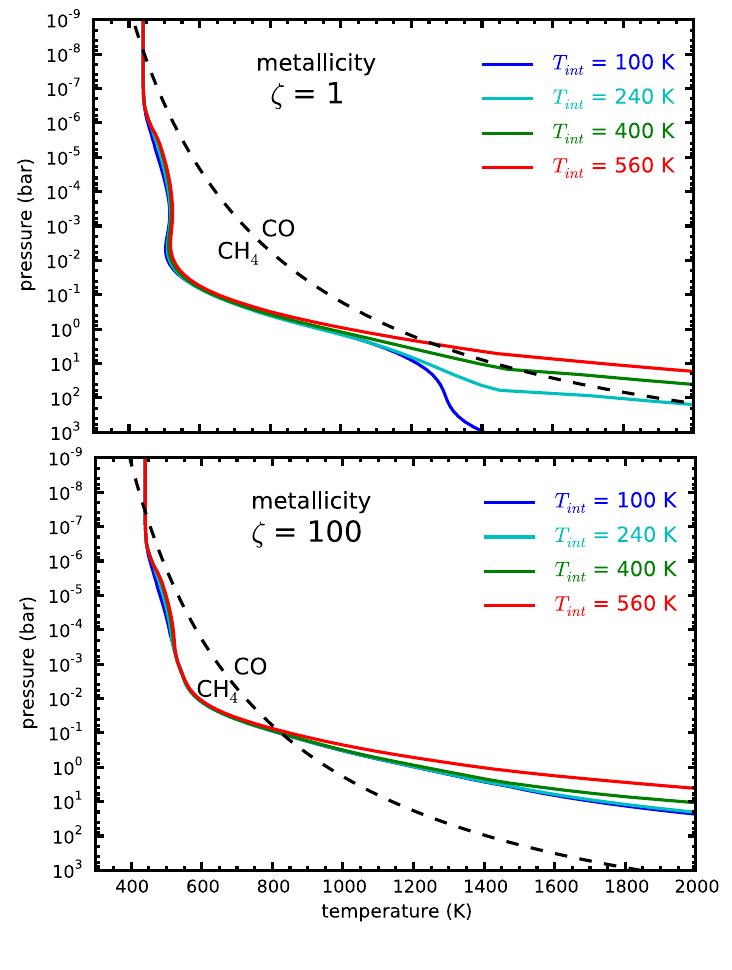}%
 \includegraphics[width=0.60\textwidth,clip]{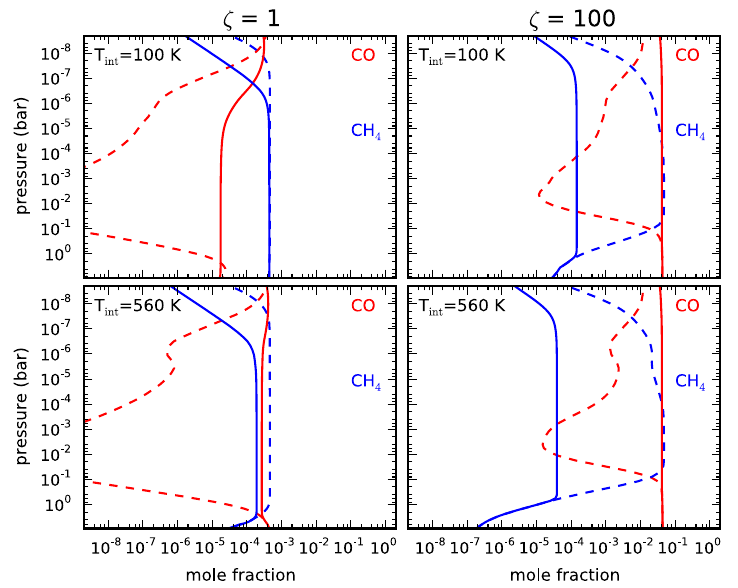}
  \caption{{\bf Left:} Thermal profiles corresponding to an atmosphere with a solar metallicity (\textit{top}) and with a metallicity 100~$\times$~solar (\textit{bottom}). Each colour corresponds to the internal temperature that varies from 100 to 560~K, corresponding to different factor of dissipation of the tidal heating. The equilibrium line CO/CH$_4$ is represented with a dashed line. {\bf Right:} Abundance profiles of CO (red) and CH$_4$ (blue) for some selected cases. The thermochemical equilibrium is represented with a dashed line and the result of the chemical model with a full line.}\label{CO_CH4_GJ436b}
\end{figure}

As you see on Fig.\ref{CO_CH4_GJ436b} (Left), the effect of the internal temperature is to heat the deep atmosphere (for a given pressure level, the hottest profile corresponds to the highest internal temperature), but the upper part of the thermal profiles ($P$ less than $\sim$ 0.1 mbar), remains quasi identical whatever is the internal temperature. We have represented the equilibrium line CO/CH$_4$, to show that depending on the internal temperature and the metallicity, the thermal profiles do not cross this line at the same level pressure, and this has a great importance when considering quenching processes. Indeed, for the high metallicity cases, quenching happens for all the $P$-$T$ profiles in a region where CO is the major C-bearing species, but for low metallicity cases, quenching can make either CO or CH$_4$ the major carbon reservoir, depending on the $P$-$T$ profiles. For the three lower internal temperatures cases ($T_{int}$=100, 240, and 400 K) quenching happens where CH$_4$ is the C-bearing species, but for the highest internal temperature case ($T_{int}$=560 K), quenching happens where CO is more abundant than methane. This is then reverberated in all the atmosphere. As you see on Fig.\ref{CO_CH4_GJ436b} (Right), except for the case $\zeta$ = 1 and $T_{int}$=100K,  even if thermochemical equilibrium predicts that methane is the C-bearing species in the middle atmosphere ([10$^{-1}$ - 10$^{-7}$] bar, area that can be probed by observations), because quenching happens deeper in the atmosphere, CO becomes the more abundant C-species in this part of the atmosphere.

We calculated synthetic spectra (see \citealt[their Figs. 4 and 5]{agu2014}) corresponding to the 12 cases we tested. Globally, their is a poor agreement with observation, but the transmission spectra measured by \cite{knu2011} seems to be consistent with models with a high metallicity and efficient tidal heating, which correspond to a methane-poor atmosphere. The relative variation of the transit depth in the 3.6 and 4.5 $\mu$m bands measured by \cite{bea2011}, which suggests a high abundance of methane, cannot be reproduced by none of our models.
The synthetic emission spectra is lower than observed during secondary eclipse, which suggests a warmer dayside atmosphere. This can be obtained by a more efficient tidal heating an/or a higher metallicity that could drive an inefficient dayÐnight heat redistribution \citep{lew2010}.

More details about this study can be found in \cite{agu2014}.

\section{Conclusions}

Using a very robust chemical scheme developed especially for high temperatures, we have studied the chemical composition of two warm Neptunes: GJ 3470b and GJ 436b. We explored the space of the unknown parameters (metallicity, temperature, vertical mixing, UV flux, tidal heating) in order to constrain their atmospheric composition. We have seen that there is a combined effect of the metallicity, the temperature, and the vertical mixing on the chemical composition of these atmospheres. Indeed, because of quenching, the composition of the middle atmosphere can be affected by temperatures found much deeper than the observations. This can explain why we deduce from observation CO-rich atmospheres, whereas chemical equilibrium predicts that CH$_4$ should dominate this kind of atmospheres. We have also found that the eccentricity of a planet have an important influence on the chemical composition of an atmosphere. Indeed, strong tidal forces generate an important internal temperature that heat the deep atmosphere, allowing, because of quenching, to have a high CO/CH$_4$ ratio.
This "carbon anomaly" depends on the temperature contrast between the probed layers, the quenching level, and the efficiency of the vertical mixing. To retrieve the elemental abundances of such atmospheres, self-consistent models that couple all these influences are needed.

\begin{acknowledgements}
O.V. acknowledges support from the KU Leuven IDO project IDO/10/2013 and from
the FWO Postdoctoral Fellowship programme. M.A., and F.S. acknowledge support from the European Research Council (ERC Grant 209622: E$_3$ARTHs). Computer time for this study was provided by the computing facilities MCIA (M\'esocentre de Calcul Intensif Aquitain) of the Universit\'e de Bordeaux and of the Universit\'e de Pau et des Pays de l'Adour.

\end{acknowledgements}

\bibliographystyle{aa}  
\bibliography{VENOT} 

\begin{thebibliography}{24}
\expandafter\ifx\csname natexlab\endcsname\relax\def\natexlab#1{#1}\fi

\bibitem[{{Ag{\'u}ndez} {et~al.}(2014){Ag{\'u}ndez}, {Venot}, {Selsis}, \&
  {Iro}}]{agu2014}
{Ag{\'u}ndez}, M., {Venot}, O., {Selsis}, F., \& {Iro}, N. 2014, \apj, 781, 68

\bibitem[{{Beaulieu} {et~al.}(2011){Beaulieu}, {Tinetti}, {Kipping}, {Ribas},
  {Barber}, {Cho}, {Polichtchouk}, {Tennyson}, {Yurchenko}, {Griffith},
  {Batista}, {Waldmann}, {Miller}, {Carey}, {Mousis}, {Fossey}, \&
  {Aylward}}]{bea2011}
{Beaulieu}, J.-P., {Tinetti}, G., {Kipping}, D.~M., {et~al.} 2011, \apj, 731,
  16

\bibitem[{{Bonfils} {et~al.}(2012){Bonfils}, {Gillon}, {Udry}, {Armstrong},
  {Bouchy}, {Delfosse}, {Forveille}, {Fumel}, {Jehin}, {Lendl}, {Lovis},
  {Mayor}, {McCormac}, {Neves}, {Pepe}, {Perrier}, {Pollaco}, {Queloz}, \&
  {Santos}}]{bon2012}
{Bonfils}, X., {Gillon}, M., {Udry}, S., {et~al.} 2012, \aap, 546, A27

\bibitem[{Elkins-Tanton \& Seager(2008)}]{elk2008}
Elkins-Tanton, L.~T. \& Seager, S. 2008, \apj, 685, 1237

\bibitem[{Knutson {et~al.}(2011)Knutson, Madhusudhan, Cowan, Christiansen,
  Agol, Deming, D{\'e}sert, Charbonneau, Henry, Homeier, {et~al.}}]{knu2011}
Knutson, H.~A., Madhusudhan, N., Cowan, N.~B., {et~al.} 2011, \apj, 735, 27

\bibitem[{Lewis \& Fegley~Jr(1984)}]{lewis1984vertical}
Lewis, J.~S. \& Fegley~Jr, M.~B. 1984, Space science reviews, 39, 163

\bibitem[{Lewis {et~al.}(2010)Lewis, Showman, Fortney, Marley, Freedman, \&
  Lodders}]{lew2010}
Lewis, N.~K., Showman, A.~P., Fortney, J.~J., {et~al.} 2010, \apj, 720, 344

\bibitem[{Liang {et~al.}(2003)Liang, Parkinson, Lee, Yung, \&
  Seager}]{liang2003}
Liang, M., Parkinson, C., Lee, A., Yung, Y., \& Seager, S. 2003, \apjl, 596,
  L247

\bibitem[{Liang {et~al.}(2004)Liang, Seager, Parkinson, Lee, \&
  Yung}]{liang2004}
Liang, M., Seager, S., Parkinson, C., Lee, A., \& Yung, Y. 2004, \apjl, 605,
  L61

\bibitem[{Line {et~al.}(2010)Line, Liang, \& Yung}]{line2010}
Line, M., Liang, M., \& Yung, Y. 2010, \apj, 717, 496

\bibitem[{Line {et~al.}(2011)Line, Vasisht, Chen, Angerhausen, \&
  Yung}]{line2011}
Line, M., Vasisht, G., Chen, P., Angerhausen, D., \& Yung, Y. 2011, \apj, 738,
  32

\bibitem[{Madhusudhan \& Seager(2011)}]{mad2011}
Madhusudhan, N. \& Seager, S. 2011, \apj, 729, 41

\bibitem[{Miller-Ricci~Kempton {et~al.}(2012)Miller-Ricci~Kempton, Zahnle, \&
  Fortney}]{mil2012}
Miller-Ricci~Kempton, E., Zahnle, K., \& Fortney, J.~J. 2012, \apj, 745, 3

\bibitem[{Moses {et~al.}(2011)Moses, Visscher, Fortney, Showman, Lewis,
  Griffith, Klippenstein, Shabram, Friedson, Marley, {et~al.}}]{moses2011}
Moses, J., Visscher, C., Fortney, J., {et~al.} 2011, \apj, 737, 15

\bibitem[{Moses {et~al.}(2013)Moses, Line, Visscher, Richardson, Nettelmann,
  Fortney, Barman, Stevenson, \& Madhusudhan}]{mos2013}
Moses, J.~I., Line, M.~R., Visscher, C., {et~al.} 2013, \apj, 777, 34

\bibitem[{Prinn \& Barshay(1977)}]{prinn1977carbon}
Prinn, R.~G. \& Barshay, S.~S. 1977, Science, 198, 1031

\bibitem[{Stevenson {et~al.}(2010)Stevenson, Harrington, Nymeyer, Madhusudhan,
  Seager, Bowman, Hardy, Deming, Rauscher, \& Lust}]{ste2010}
Stevenson, K., Harrington, J., Nymeyer, S., {et~al.} 2010, \nat, 464, 1161

\bibitem[{{Venot} {et~al.}(2014){Venot}, {Ag{\'u}ndez}, {Selsis}, {Tessenyi},
  \& {Iro}}]{venot2014}
{Venot}, O., {Ag{\'u}ndez}, M., {Selsis}, F., {Tessenyi}, M., \& {Iro}, N.
  2014, \aap, 562, A51

\bibitem[{{Venot} {et~al.}(2013){Venot}, {Fray}, {B{\'e}nilan}, {Gazeau},
  {H{\'e}brard}, {Larcher}, {Schwell}, {Dobrijevic}, \& {Selsis}}]{venot2013}
{Venot}, O., {Fray}, N., {B{\'e}nilan}, Y., {et~al.} 2013, \aap, 551, A131

\bibitem[{{Venot} {et~al.}(2012){Venot}, {H{\'e}brard}, {Ag{\'u}ndez},
  {Dobrijevic}, {Selsis}, {Hersant}, {Iro}, \& {Bounaceur}}]{venot2012}
{Venot}, O., {H{\'e}brard}, E., {Ag{\'u}ndez}, M., {et~al.} 2012, \aap, 546,
  A43

\bibitem[{Visscher \& Moses(2011)}]{vis2011}
Visscher, C. \& Moses, J.~I. 2011, \apj, 738, 72

\bibitem[{Wakelam {et~al.}(2012)Wakelam, Herbst, Loison, Smith, Chandrasekaran,
  Pavone, Adams, Bacchus-Montabonel, Bergeat, Beroff, Bierbaum, Chabot,
  Dalgarno, van Dishoeck, Faure, Geppert, Gerlich, Galli, Hebrard, Hersant,
  Hickson, Honvault, Klippenstein, Le~Picard, Nyman, Pernot, Schlemmer, Selsis,
  Sims, Talbi, Tennyson, Troe, Wester, \& Wiesenfeld}]{wakelam_kida2012}
Wakelam, V., Herbst, E., Loison, J.-C., {et~al.} 2012, The Astrophysical
  Journal Supplement Series, 199, 21

\bibitem[{Zahnle {et~al.}(2009{\natexlab{a}})Zahnle, Marley, Freedman, Lodders,
  \& Fortney}]{zahnle2009a}
Zahnle, K., Marley, M., Freedman, R., Lodders, K., \& Fortney, J.
  2009{\natexlab{a}}, The Astrophysical Journal Letters, 701, L20

\bibitem[{Zahnle {et~al.}(2009{\natexlab{b}})Zahnle, Marley, \&
  Fortney}]{zahnle2009b}
Zahnle, K., Marley, M.~S., \& Fortney, J.~J. 2009{\natexlab{b}}, eprint
  arXiv:0911.0728

\end{thebibliography}

\end{document}